\documentclass[preprint,sort&compress]{elsarticle}

\usepackage{lineno,hyperref}
\usepackage[english]{babel}
\usepackage[latin1]{inputenc}
\usepackage{graphicx}
\usepackage{amssymb, amsmath}
\usepackage[usenames,dvipsnames]{color}

\journal{Nuclear Instruments and Methods A}


\usepackage{xspace}

\usepackage{listings}

\begin{document}
\def \Xmax {X$_{\rm max}$}
\def \Smu {S$_\mu$}

\begin{frontmatter}

%\title{Elsevier \LaTeX\ template\tnoteref{mytitlenote}}
\title{Layered water Cherenkov detector for the study of ultra high
  energy cosmic rays}

\author[lpnhe]{Antoine Letessier-Selvon\corref{cor1}}
\ead{antoine.letessier-selvon@in2p3.fr}
\cortext[cor1]{Corresponding author}

\author[lpnhe]{Pierre Billoir}
\author[lpnhe]{Miguel Blanco}
\author[lpnhe,ugr]{\\Ioana C. Mari\c{s}} 
\author[lpnhe]{Mariangela Settimo}

\address[lpnhe]{LPNHE, UPMC University Paris 6, UPD University Paris 7, CNRS/IN2P3, 4 place Jussieu, FR-75252 Paris, France.}
\address[ugr]{University of Granada and C.A.F.P.E., Cuesta del Hospicio, 18071, Granada, Spain}

\begin{abstract}
We present a new design for the water Cherenkov detectors that are in
use in various cosmic ray observatories. This novel design can provide a significant improvement in
the independent measurement of the muonic and
electromagnetic component of extensive air showers. From such 
multi-component data an event by event classification of the primary
cosmic ray mass becomes possible. According to popular hadronic
interaction models, such as EPOS-LHC or QGSJetII-04, the
discriminating power between iron and hydrogen primaries reaches
Fisher values of $\sim$ 2 or above for energies in excess of
$10^{19}$~eV with a detector array layout similar to that of the Pierre
Auger Observatory.
\end{abstract}

\begin{keyword}
\texttt{UHECR}\sep Mass composition \sep Surface array \sep Cherenkov detectors
\end{keyword}

\end{frontmatter}

\section{Introduction}
At the highest energies, above 10$^{19}$~eV (10 EeV or 1.6 Joules),
cosmic rays are scarce (less than one per km$^2$ per year) and with
limited data on their nature.  This situation dramatically limits the
contribution of Ultra High Energy Cosmic Ray (UHECR) physics to
fundamental physics.  However, if one could access the nature of the
primary particle and the details of the cascade evolution and content,
substantial information would be collected about hadronic interactions
above 100 TeV center-of-mass and about the sources of such energetic
particles. Aiming at an excellent primary cosmic ray identification,
multi-parametric measurements of Extensive Air Showers (EAS) have the
potential of measuring hadronic cross sections above 100 TeV and up to
450 TeV center-of-mass, constraining interaction models, detecting or
setting limits on UHE neutrino or gamma ray fluxes, identifying UHECR
sources, and constraining Galactic and intergalactic magnetic fields.

Such a collection of data regarding particle interactions and cascade
development at the highest energies will certainly play an invaluable
role in our understanding of fundamental physics (and regarding the
nature and content of our Universe).  However, the limitations of
current measurements regarding, in particular, the nature (or
composition) of the cosmic ray flux above 50 EeV strongly limit the
coherent understanding of the available data and our capacity to
address fundamental questions regarding particle interactions and
transport at the highest energies, together with the nature of the
sources of the particles. Even today, the existence of the
Greisen-Zatsepin-Kuz'min (GZK)
cut-off~\cite{Greisen:1966jv,Zatsepin:1966jv} is still not totally
resolved as the observed reduction in the cosmic-rays spectrum can
still be interpreted as the sources ``running out of steam''.

From the experimental point of view it is clear that large statistics
and high-precision composition measurements at the highest energy
(above a few tens of EeV) are needed. The current
results~\cite{Abbasi:2009nf,Abraham:2010yv,Barcikowski:2013nfa,Array:2013dra,Dedenko:2012ws}
show that these measurements can hardly be made, with the required
statistics, using current setups, in particular, because of the low
duty cycle of current fluorescence
detectors~\cite{Abraham:2009pm,Abreu:2010zzl,Tokuno:2012mi}.

In this paper, we introduce a novel design of water Cherenkov
detectors which allows one to measure independently the muonic and
electromagnetic components of the EAS and will not suffer of the duty cycle
limitations of the fluorescence detection technique. The
design concept and its application to water Cherenkov stations as in
the Pierre Auger Observatory (Auger) are described in
section~\ref{sect:Design}. Its properties and performance for mass
composition studies are discussed in
section~\ref{sect:Performance}. A first prototype has been
successfully installed at the Auger site and first results are shown
in section~\ref{sect:Prototype}.

 \begin{figure}[!t]
\center \includegraphics[width=0.8\textwidth]{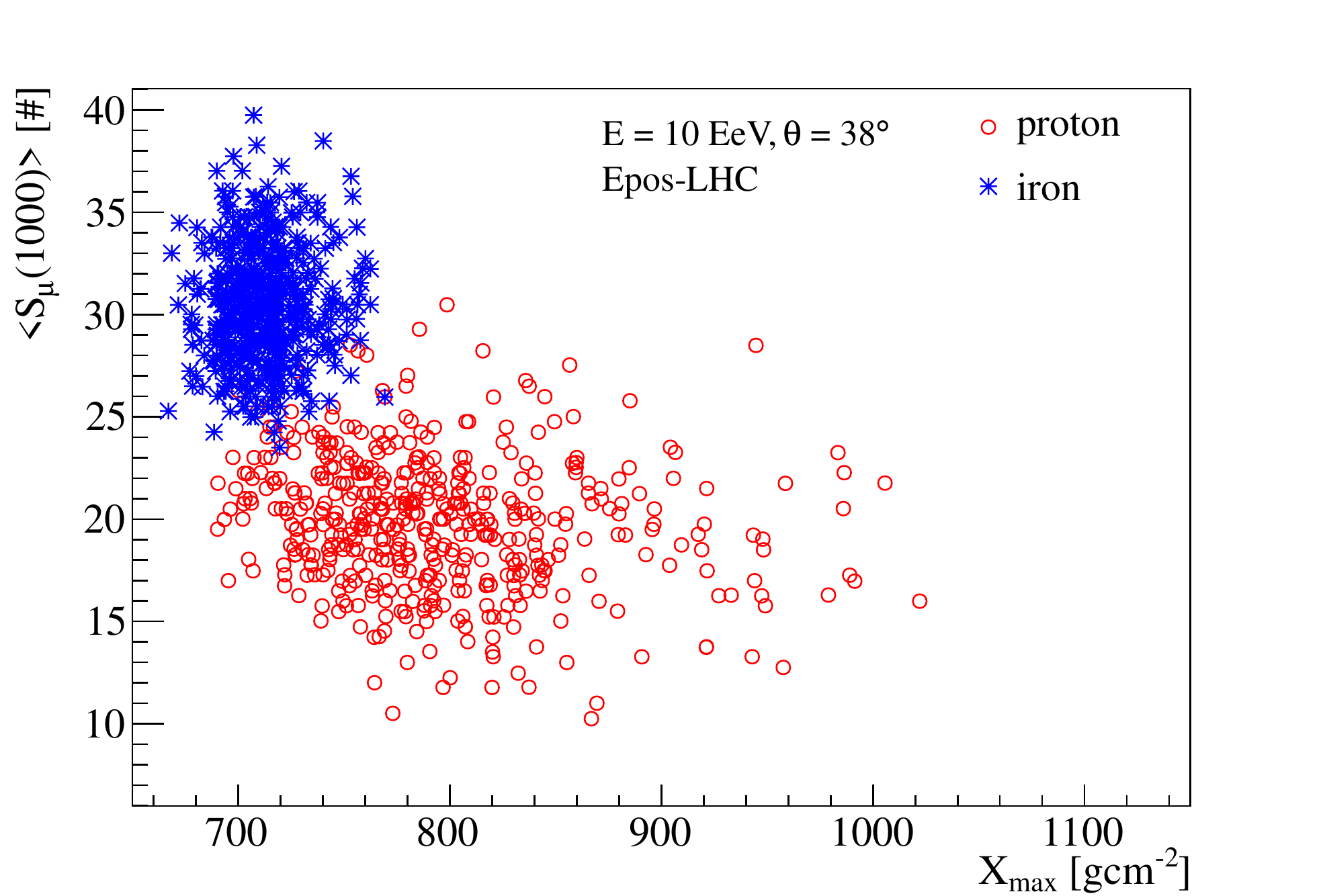}
  \caption{Average number of muons, $<$ \Smu\,(1000)$>$,  entering a 10~m$^2$ detector versus \Xmax\, for 500 iron (blue) or proton (red) showers at 10 EeV and 38$^\circ$ zenith angle. The value for \Smu\, corresponds to the average of four measurements. The \Xmax\, distribution (horizontal spread)  is due to the shower to shower development fluctuations for a given primary mass at fixed zenith angle and energy.}
  \label{fig1}
 \end{figure}

\section{Design principles}\label{sect:Design}

\subsection{Why count muons}

After the few first interactions, EAS generated by UHECR contain a
very large number of particles undergoing an even larger number of
interactions. One can therefore expect that the cascade description
can be modeled from a reduced set of universal functions that will
depend on the shower age and whose relative amplitudes will carry the
information on the primary type and energy.  This ``universality''
concept~\cite{Lipari:2009zz,Ave-ICRC2011} has been extensively studied
and validated with Monte Carlo simulations and, indeed, the arrival
direction, the depth of shower maximum (\Xmax\,) and the muon size at
ground (\Smu\,) effectively contain the essential information about
the primary UHECR.  It follows that one can parametrize the ground
signal of EAS with a set of universal functions (predicted or measured
and only dependent on those macroscopic quantities), from which the
relevant information about the primary UHECR can be
retrieved~\cite{Ave-ICRC2011}.

The principle of cosmic ray mass determination using the muon
densities at ground and the shower maximum is illustrated in
figure~\ref{fig1} where a set of 10 EeV proton and iron simulated
showers at 38 degrees zenith angle has been used. It is clear that,
once the energy is known, a selection criterion can be built in the
\Xmax\, vs \Smu\, plane that permits a very high efficiency/purity
separation between the two species.  Hence, a detector design that can
perform an appropriate measurement of both these quantities (or
related ones) should provide the necessary information to analyse cosmic ray
data within definite mass samples. In figure~\ref{fig1} no
detector resolution is taken into account. In the next sections the
impact of a more realistic situation is presented. However, it is
evident from this figure that, due to shower to shower fluctuations, a
resolution of about 20~g/cm$^2$ for \Xmax\, is needed to infer the
mass composition on the event by event basis. On the other hand, a
resolution of about 10 to 20\% for \Smu\, is also suitable given 
Poisson fluctuations in the muon numbers at ground and the limited
surface area of muon counters that can be realistically deployed
($\sim$~10~m$^2$).

\subsection{Layered system}

Energy absorption in water Cherenkov detectors (WCD) can be used to
distinguish muons from the electromagnetic (EM) component of EAS.  The
electromagnetic particles (e$^{+}$/e$^{-}$ and photons) reaching the
ground have an average energy of about 10 MeV. Electrons and positrons
are absorbed in water over a few centimeters while photons deposit
 their energy over a radiation length ($\sim$~36~cm). On the
other hand muons, having a typical average energy of a few GeV, can go
through several meters of water without being stopped. If the water volume of a
WCD is split into two horizontal layers, one of which being at least
partially shielded from the electromagnetic component, then the two
volumes will provide distinct responses to the EAS components.  A
Layered Surface Detector (LSD) can therefore provide the means to
reconstruct independently, for each EAS, the muonic and
electromagnetic components. One remarkable property is that the
shielding of the muon sensitive volume does not need to be very large:
in practice a single radiation length is enough to achieve good
performance. This is due to the fact that WCD have a distinct
response to the EM and muonic component of EAS. Indeed, whereas EM
particles of the EAS are absorbed in the water volume in a
calorimetric way, muons are minimum ionizing particles and deposit an
amount of energy proportional to their track length (i.e., several
hundreds of MeV for a meter depth detector).

\begin{figure}[!t]
\begin{center}
  \includegraphics[width=\columnwidth]{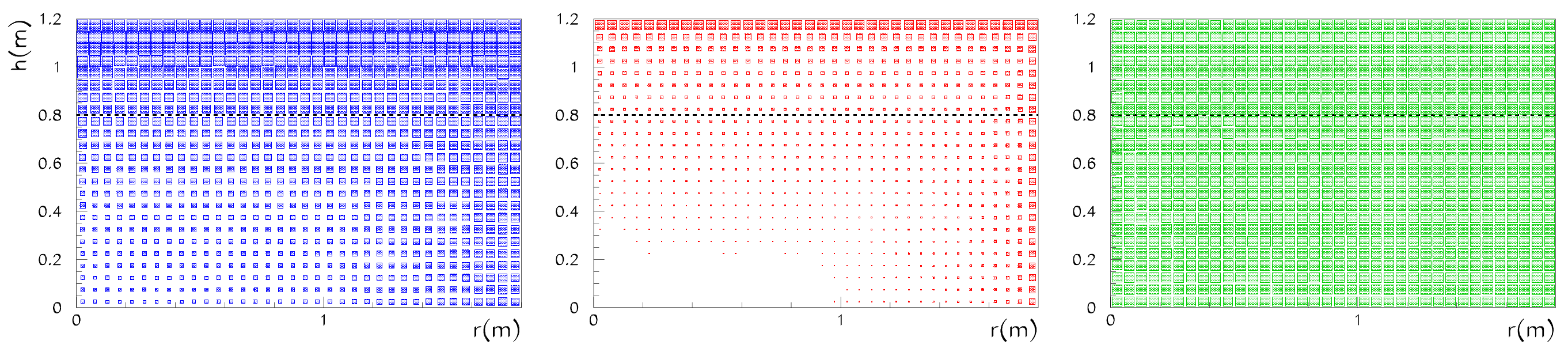}
  \end{center}
  \caption{Distribution of the Cherenkov photons production point in a
    1.2~m height and 1.8~m radius WCD. From left to right the
    contribution from the photons, $e^+e^-$ and muon component of a
    30~EeV EAS with 45$^o$ zenith angle is shown.}
  \label{fig2}
 \end{figure}

In its simplest form, a LSD is composed of two independent and
light-tight volumes that are created by inserting a horizontal
reflective layer in a water volume.  Other implementations with more
than two layers and/or shielding on the side can also be imagined. 
However the simple design described here is enough to show the
main properties of such a detector. In the next section, a
modification of the WCD design from the Auger surface array~\cite{AugerNIM}
is presented based on this design.

\subsection{A LSD design from the Auger WCDs}

The Auger WCD~\cite{Allekotte:2007sf} have a 1.8\,m radius by 1.2\,m
height water volume, overlooked by three 9 inch photomultiplier tubes.
In figure~\ref{fig2} we show the production point of Cherenkov photons
in such detectors, for a 30~EeV shower with 45$^\circ$ zenith
angle. As expected the EM component dominantly deposits its energy on
the top and side of the station.

The two water volumes of the LSD can be created by inserting a
horizontal reflective layer at a height of 80~cm from the WCD
bottom\footnote{This choice is justified in the next paragraph}. This
is represented as a dashed line in figure~\ref{fig2}. The three PMTs
that equip the Auger WCD are left in place.  In the upper layer, the EM
component of an EAS will deposit most of its energy, while the 80~cm
bottom layer will detect the much more penetrating muons (and to a
lesser degree the surviving EM particles).

\begin{figure}[!t]
\begin{center}
  \includegraphics[width=\columnwidth]{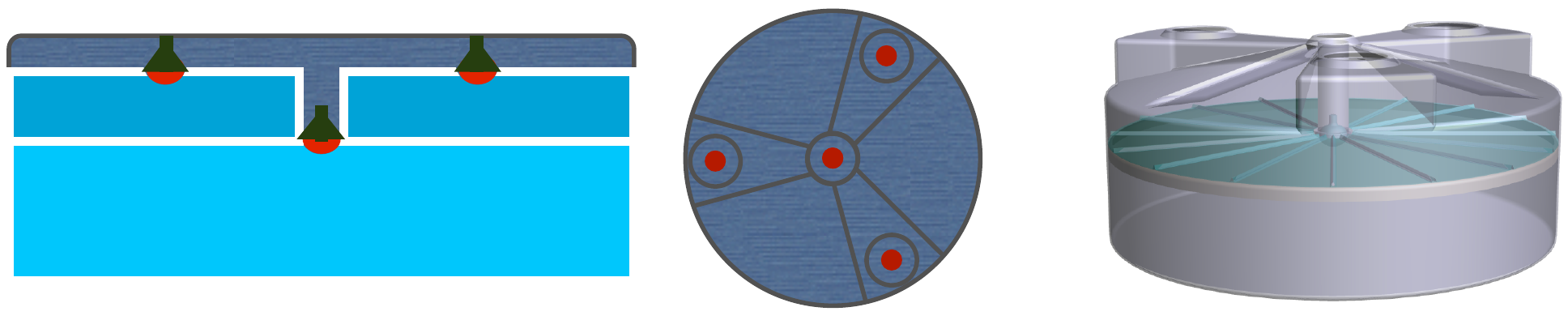}
  \end{center}
  \caption{Schematic and artistic view of a LSD build from an Auger WCD design.}
  \label{fig3}
 \end{figure}

A central cylinder of about 10 inch diameter provides the mechanical
structure as well as the enclosure for the additional 9 inch
photo-tube that collects the light from the bottom layer. Some
simplified schematics and an artistic view of this LSD is shown in
figure~\ref{fig3}.

\subsection{Signal extraction}

The reconstruction of the EM and muonic component of EAS relies on the
fraction of the signal deposited by each component in each of the two
layers. These fractions define a $2\,\times\,2$ matrix $\mathcal{M}$
that gives the measured top and bottom signal ($S_\text{top}$ and
$S_\text{bot}$ respectively) as a linear superposition of the EM
($S_\text{EM}$) and muon ($S_\mu$) contributions (the column sums of
the matrix are one by construction),
\begin{equation}
\begin{pmatrix} S_\text{top} \\ S_\text{bot} \end{pmatrix} =
  {\mathcal{M}} \begin{pmatrix} S_\text{EM} \\ S_\mu\end{pmatrix} =
  \begin{pmatrix}a & b \\ 1-a & 1-b \end{pmatrix}
    \begin{pmatrix} S_\text{EM} \\ S_\mu\end{pmatrix}
\end{equation}

Hence, if the matrix $\mathcal{M}$ can be inverted, the muonic and EM
signal deposition in the LSD can be retrieved as :

\begin{equation}
    \begin{pmatrix} S_\text{EM} \\ S_\mu\end{pmatrix} = 
  {\mathcal{M}^{-1}} \begin{pmatrix} S_\text{top} \\ S_\text{bot} \end{pmatrix}
\end{equation}

The determinant $\mathcal{D}$ of the $\mathcal{M}$ matrix is $a-b$ and
maximises when $\mathcal{M}$ is equal to the identity ($a=1$ and
$b=0$). In a realistic situation $a$ is always less than one while $b$
is always larger than zero, hence $\mathcal{|D|}$ will be less than
one. This is important as the statistical uncertainty in the
reconstructed muonic and EM signals from measurements in the top
and bottom layer are driven by $1/\mathcal{D}$, the determinant of
$\mathcal{M}^{-1}$.

The coefficients $a$ and $b$ depend on the geometry of the two water
volumes and on the efficiency of the light collection. They can be
obtained from well established simulations of the tank response.
A rough estimate of $a$ and $b$ can be derived for a vertical
incidence, when neglecting the signal produced by particles entering
through the side of the detector. In fact, modelling the absorption of
the electromagnetic component by an exponential decay according to the
radiation length $X_0$ and for a tank of height $H$ with a layer
interface located at a distance $H-h$ from the bottom we have :
\begin{equation}
a = 1-e^{-h/X_0} \mbox{    and   } b = \frac{h}{H} \mbox{   with   } h\in[0,H] 
\end{equation}
 $\mathcal{D} = a-b$ is maximum for $h = X_0 \ln(H/X_0)$. If $H$ is
large enough (keeping the radius also large so that the side
contributions can still be neglected) $a$ goes to 1 while $b$ goes to
0 and $\mathcal{M}$ tends towards the ideal unity matrix. In a more
realistic case, particles entering from the side wall of the
station cannot be neglected and the optimal values of $a$ and $b$
also depend on the proportion between the height and the radius of the
tank.

In the particular case of the Auger WCD, $H$ is 120~cm, $X_0=36$~cm
giving $h=43$~cm. An optimal position for the interface layer is
therefore at about 80~cm from the bottom of the water tank.

\section{Performances}\label{sect:Performance}
\subsection{Matrix universality}

To precisely calculate $a$ and $b$ and to characterize the
performances of the LSD, 
simulations of the detector response have been performed.
Air showers
have been simulated with the CORSIKA code~\cite{Heck98a}, using
EPOS-LHC~\cite{PhysRevLett.106.122004} and
QGSJetII.04~\cite{Ostapchenko:2010vb} as high energy interaction
models and FLUKA~\cite{Ferrari:2005zk} at low energy. Various
libraries have been generated with a uniform distribution in
$\cos^2\theta$ for different primary type (proton, helium, nitrogen
and iron) and in two energy intervals (from 8 to 13 EeV and from 40 to
60 EeV, uniformly distributed in the logarithm of energy).

\begin{figure}[t]
\begin{center}
\includegraphics[width=0.49\textwidth]{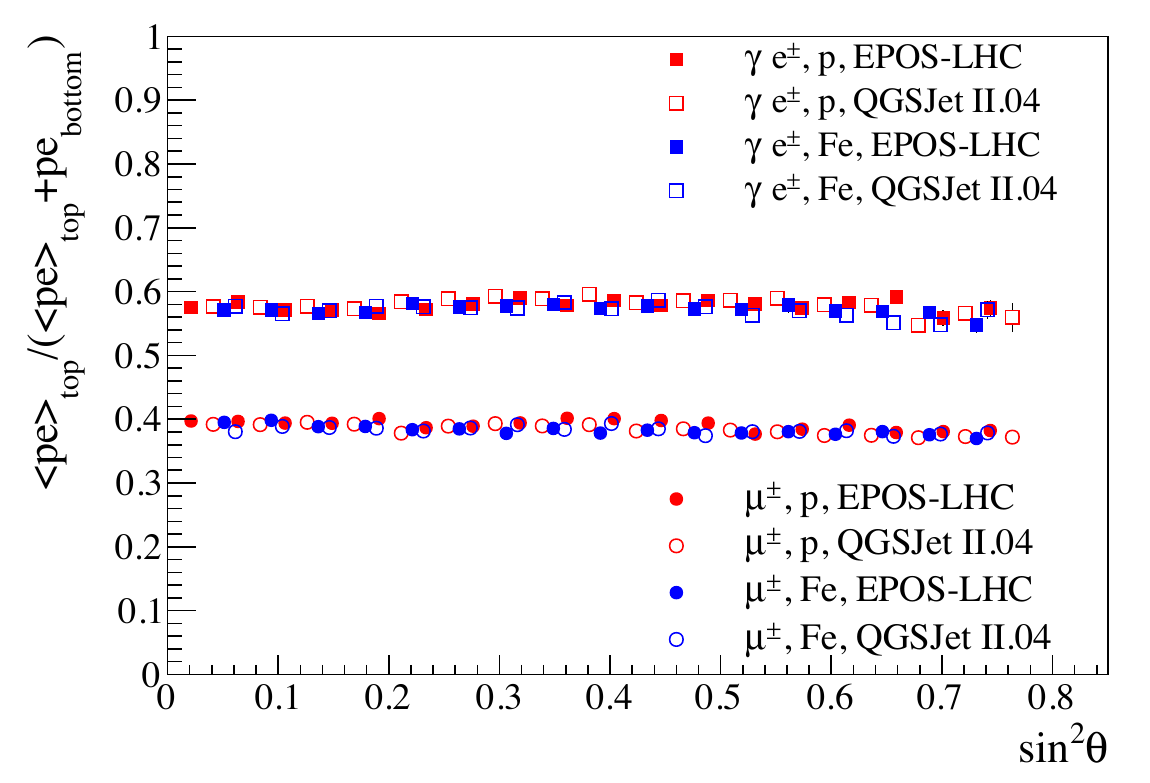}
\hfill
\includegraphics[width=0.49\textwidth]{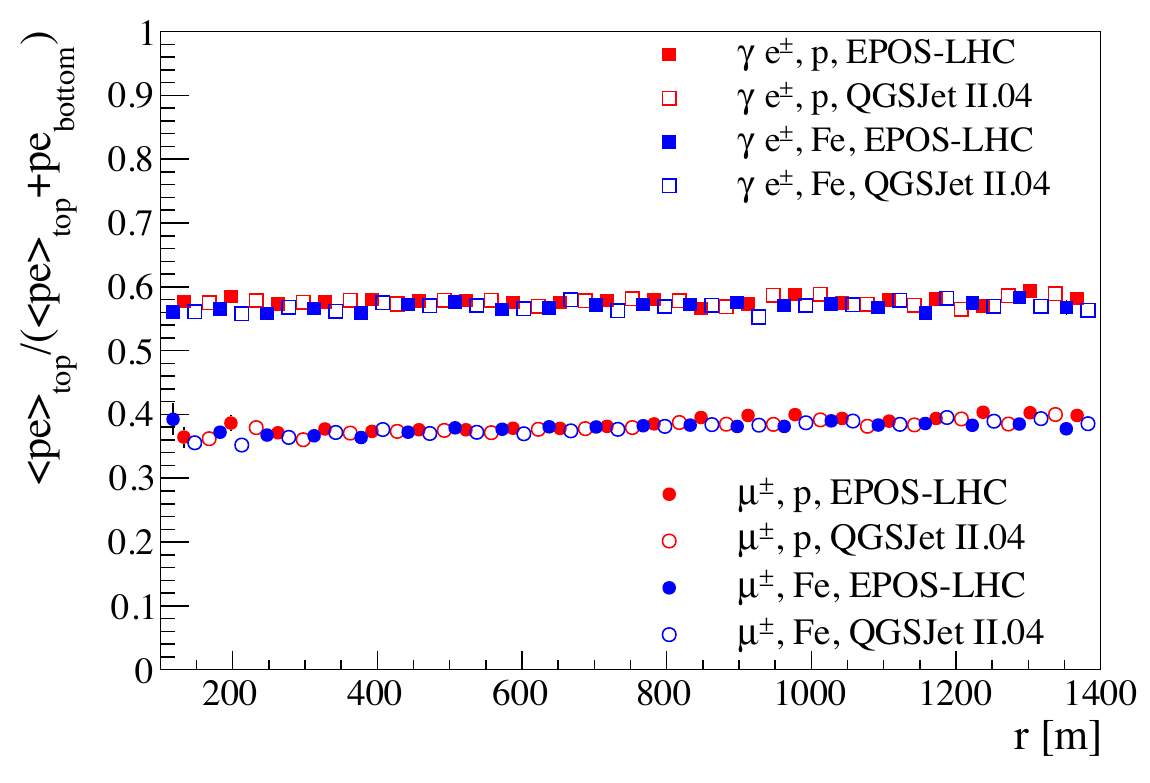}
\end{center}
\caption{Fraction of photo-electrons (pe) collected in the top layer
  for the EM (square symbols, $a$ coefficient of the matrix) and
  muonic (round symbols, $b$ coefficient) component for two hadronic models.  Left :
  dependence as a function of zenith angle. Right: dependence as a
  function of distance to the shower core. The coefficients $a$ and
  $b$ are essentially independent of the shower characteristics or
  detector distance from the shower core, they only depend on the tank
  geometry.}
\label{figab}
\end{figure}

\begin{figure}[t]
\begin{center}
\includegraphics[width=0.95\textwidth]{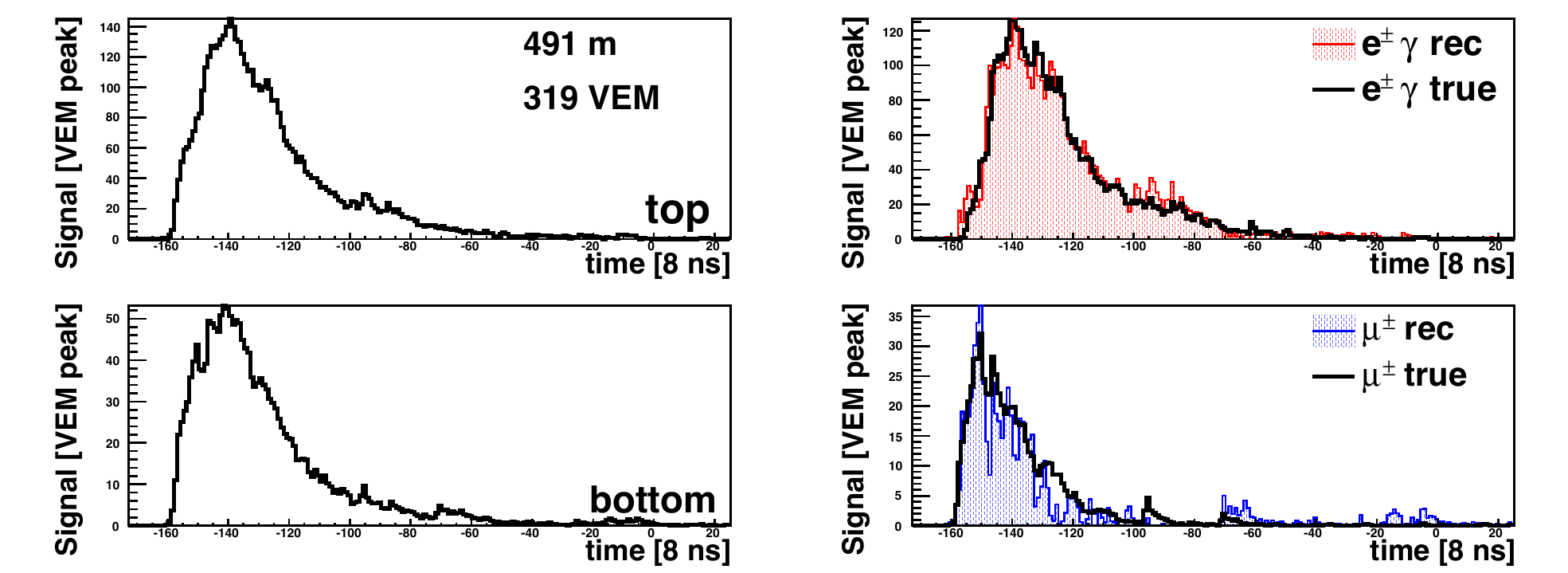}
\end{center}
\caption{Simulation of the signal collected in the top (bottom)
  part of a 10\,m$^2$ LSD for a 11 EeV shower with 46$^\circ$ zenith angle and
  491 meters away from the core. The $\mathcal{M}$ reconstruction of
  the muon and EM traces compared to the generated ones are shown in
  the right panel. The agreement is striking both in shape (timing
  information of the two components) and amplitude. This hints at the
  excellent performances that can be expected from this design for the
  multi-component study of EAS.}
\label{figemu}
\end{figure}

The matrix coefficients $a$ and $b$ are derived from simulations as
the ratio between the photo-electrons collected in the top layer and
the total number of photo-electrons in the two volumes for the EM and
muonic components respectively.  A remarkable property of the LSD is
that the coefficients $a$ and $b$ are essentially independent of the
UHECR primary type and energy and also of the particular simulation
model used to describe the EAS. This is shown in figure~\ref{figab}
where the coefficient $a$ and $b$ have been evaluated for different
primaries and hadronic models and are shown as a function of zenith
angle and distance to core. It is notable that for the Auger WCD
geometry, $a$ and $b$ are also essentially independent of the shower
zenith angle in the range [0, 60$^\circ$].  This is due to a
compensation between the top and side wall contributions coming from
the particular geometry of the Auger WCD which have an height over
radius ratio of 2/3.

For the particular case of the LSD from the modified Auger WCD the
parameter $a$ is nearly 0.6 while $b$ is about 0.4 (see
figure~\ref{figab}), leading to a determinant $\mathcal{D}$=1/5.

\subsection{Signal reconstruction}

For ground arrays the reconstruction of the primary UHECR properties
relies on the adjustment of a lateral distribution function (LDF) that
describes the detector signals as a function of their distance from
the core. The value of the LDF at a reference distance to core\footnote{The reference distance is chosen as the one which
  minimizes the fluctuations of the expected signal, due to the lack
  of knowledge on the LDF.  It is a function of the array grid spacing
  and the EAS energy range studied. It is 1000~m for
  Auger~\cite{Newton:2006wy}.} serves as an energy estimator but other
characteristics such as its slope also reflect EAS properties, for
instance its age. Using the LSD, it is possible to derive two
independent LDFs, one for the muonic and one for EM component, by
means of the corresponding signals reconstructed in each station. This
offers the additional possibility to perform an energy reconstruction
that takes into account the muon size and shower age and to produce 
a calibration based only on the electromagnetic component. Such a procedure 
would be less sensitive  to the shower to shower fluctuation and to the
interaction models than the one that uses the mixed EM and muonic signals. 

\begin{figure}[t]
\includegraphics[width=0.485\textwidth]{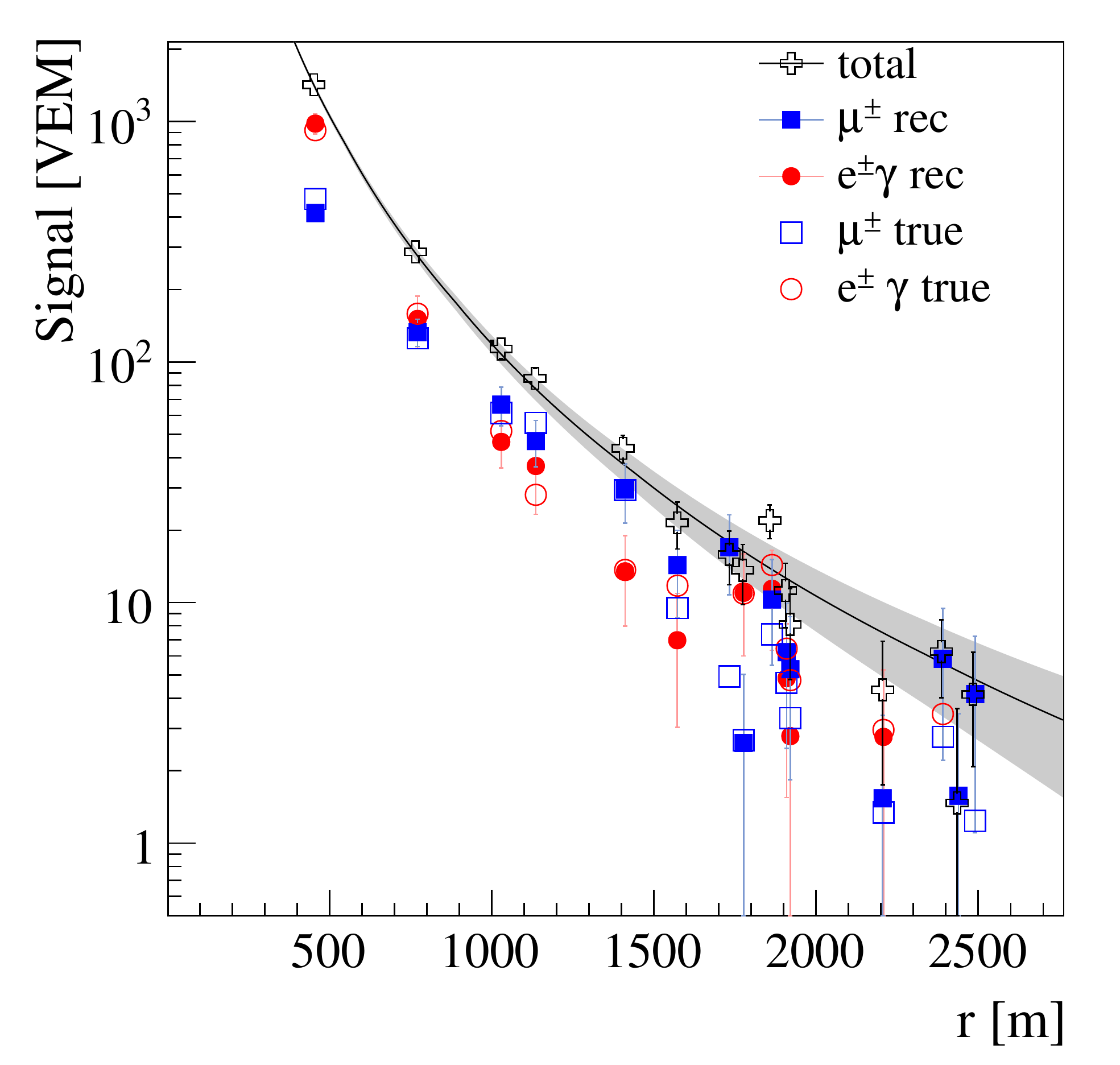}
\hfill
\includegraphics[width=0.49\textwidth]{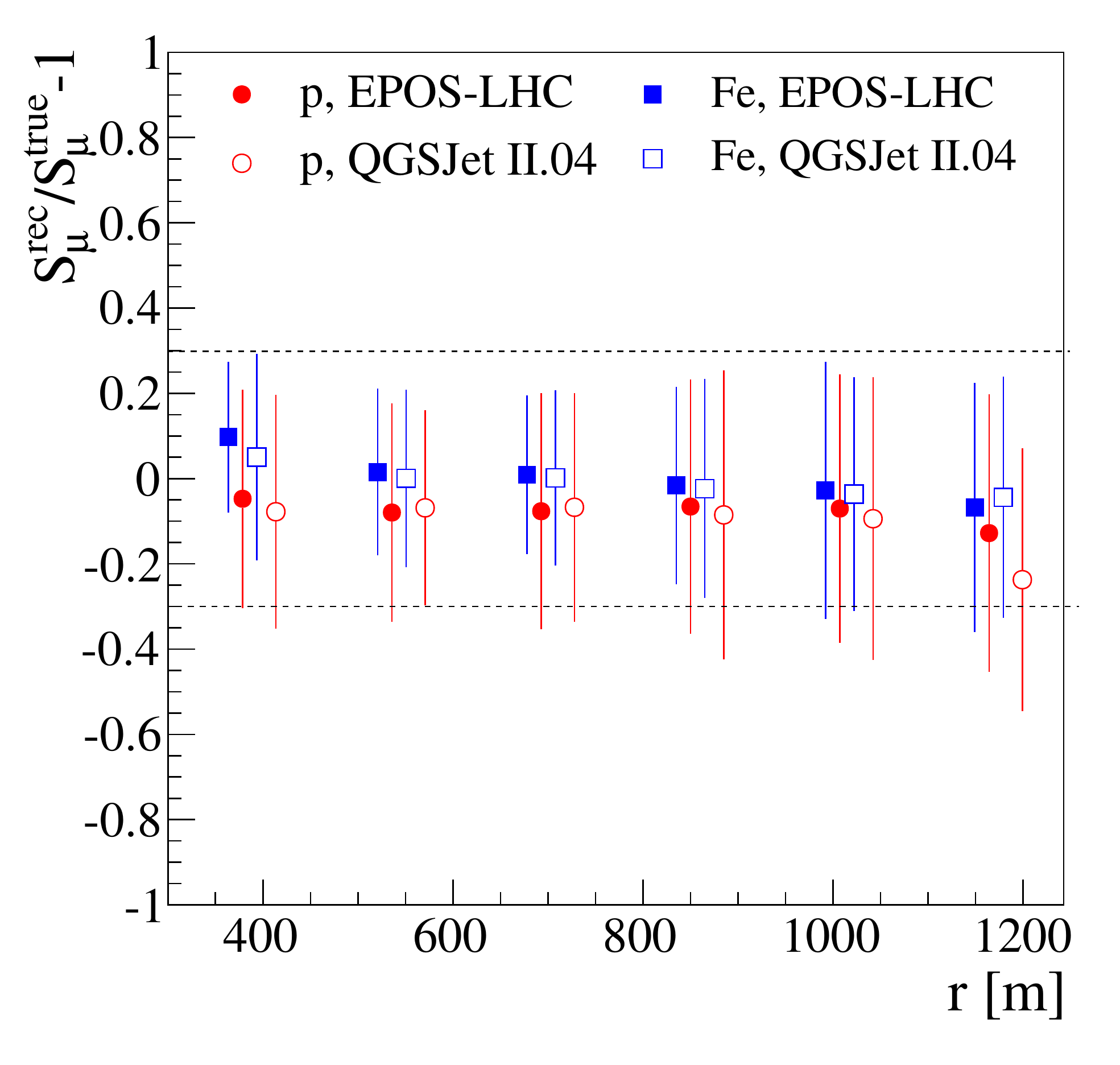}
\caption{Left: example of muonic and EM
  independent LDFs reconstructed using the individual signal
  reconstruction in each of the detectors. 
 Right: individual LSD muon signal reconstruction as a function of the distance to core, 
the error bars represent the signal resolution.
}
\label{LSD-3}
\end{figure}

Inverting the matrix $\mathcal{M}$, it is possible to
construct for each detector FADC traces for the two components
separately. An example is shown in figure~\ref{figemu} for an 11 EeV shower at 46$^\circ$ zenith angle. This
graph alone demonstrates the power of the LSD to accurately determine
the muonic and electromagnetic components of the EAS, for both the
integrated signals and their time distribution. The muonic and
electromagnetic LDFs can easily be reconstructed from these individual
measurements, as shown in figure~\ref{LSD-3} (left).

\begin{figure}
\includegraphics[width=0.505\textwidth]{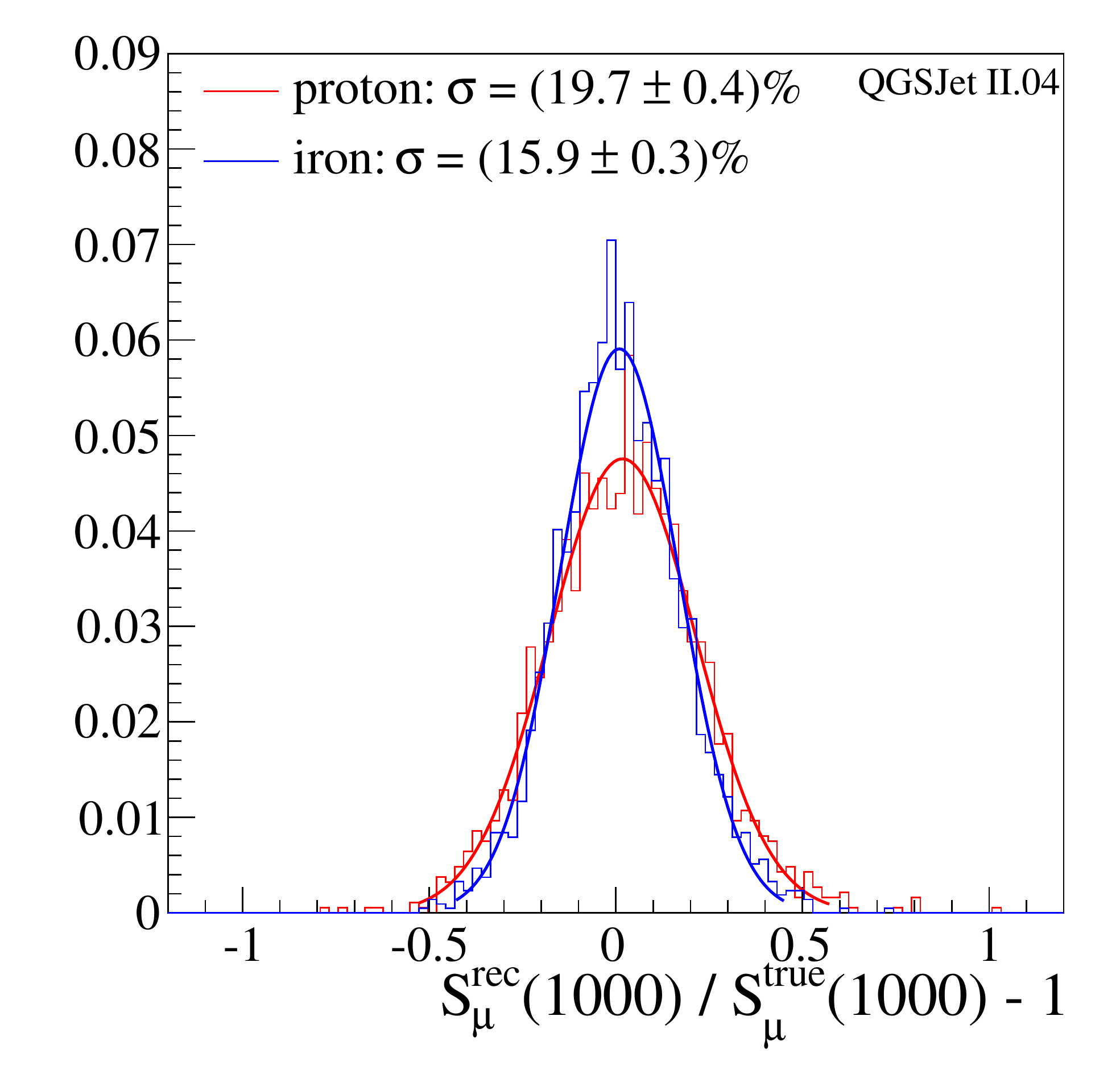}
\hfill
\includegraphics[width=0.49\textwidth]{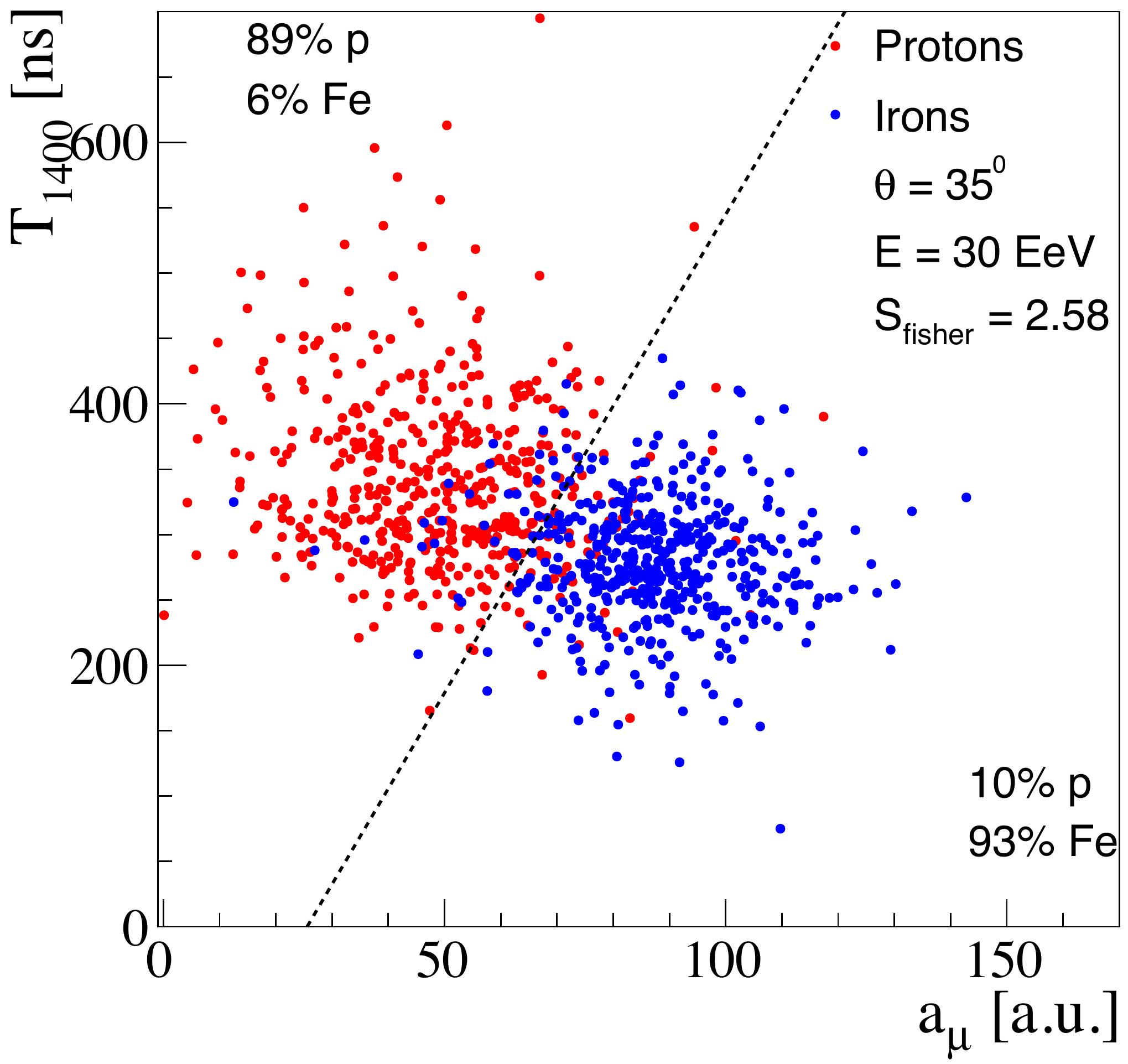}
\caption{ Left: global resolution on the muon size at 1000~m
  $S_{\mu}$ for air showers with energy between 8 and 13 EeV.
  Right: Example of proton / iron separation plot at 30\,EeV from the LSD reconstructed
  start time and muon signal size. The horizontal axis is the muon
  size at 1000~m from the shower core in arbitrary units, it is
  obtained by a simple LDF fit to the reconstructed muon signal in
  each WCD. The vertical axis is the start time at 1400\,m (see
  text).}
\label{LSD-3b}
\end{figure}
 \subsection{Mass separation}

With the LSD one can reconstruct the muon signals from at distances to core
of nearly 200~m to more than 2\,km for the highest energetic showers. 
As shown in figure~\ref{LSD-3} (right), the signal resolution in each
detector is better than 25\% when more than 20 muons enter the
detector while the Poisson fluctuations dominate for smaller
signals. These resolutions are given for 10~EeV and they improve
significantly above 40 EeV, e.g.\ ${<}$14\% for the muon size.

By following the usual approach for the shower size reconstruction
with surface array detectors, a measurement of the muon size of EAS
can be obtained from the muonic LDF, at a reference
distance\footnote{We adopt here the same reference distance as in
  Auger (1000~m), but optimal values can be derived in the future.}.
The global resolution on the EAS muon size parameter is 20\% (16\%)
for proton (iron) at 10 EeV, improving to about 10\% for both at 70 EeV
(see figure~\ref{LSD-3b}, left).

The muon size can be combined with \Xmax\, or with other age sensitive parameters to produce a two or multi-dimensional plot as discussed in the introduction, improving the mass composition separation capabilities. 
An \Xmax\, sensitive parameter can also be retrieved following the
universality principle of EAS description, as for example using the 
approach proposed in~\cite{Ave-ICRC2011,UniversalityTime}. 
An example of the separation power for a 50\%~proton 50\%~iron mixed composition at 30\,EeV is
given in figure~\ref{LSD-3b} (right). 
The muon signal reconstructed with the LSD is plotted against the signal start time at 1400~m (T$_{1400}$) which measures the delay of the shower particles with respect to the arrival time of a
imaginary planar front.  This parameter is sensitive to the shower
front curvature and correlates to \Xmax\,. In this particular example
we have used a timing resolution of 8~ns which can be easily achieved
with modern GPS receivers. This correspond to an \Xmax\, resolution of 
40 to 60\,g/cm$^2$ depending on zenith angle. 
The Fisher separation coefficient for this particular case
is larger than 2 indicating that excellent separation power. 
It is worthwhile noting that this merit factor is obtained for a fixed energy in the simulation. In a realistic scenario both, the energy and S$_\mu$, need to be estimated from the LSD data. It is to be expected that this will diminish the proton-iron separation, but since we can experimentally estimate the energy by combining the \Xmax\, (or T$_{1400}$) with the shower size of the electromagnetic component at ground, a merit factor of better than 2 will be possible.
The parameter T$_{1400}$ is given as example and it is not meant to be considered as the optimal variable for the analysis. A detailed study of \Xmax\, (or age) sensitive parameters and of the mass separation
is in progress and is however out of the scope of this paper. 

 \begin{figure}[t]
\center
\includegraphics[width=0.9\textwidth]{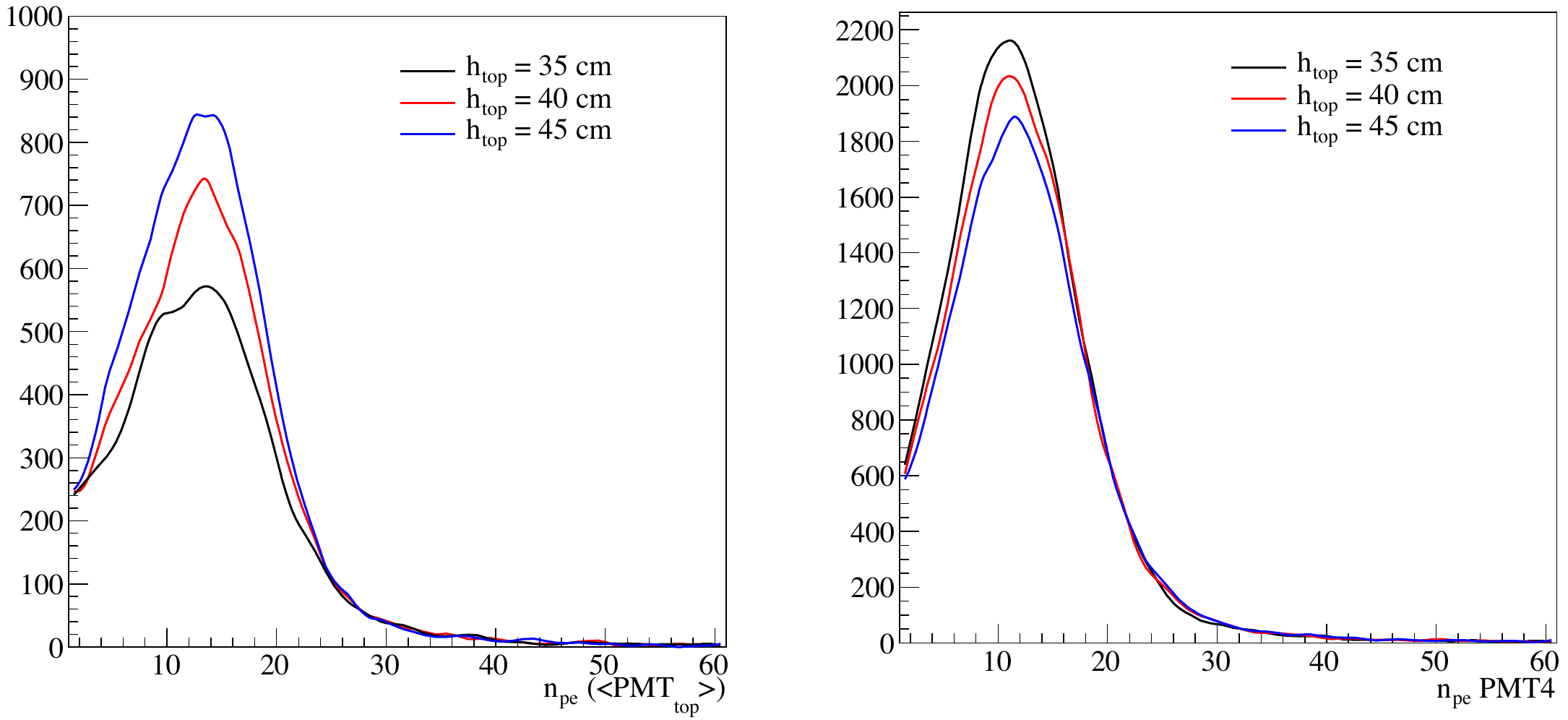}
\caption{\small Signals from the electrons of muon decaying in the LSD
  tanks. Left: average signal in the three top PMTs. Right: signal in
  the bottom PMT.  The statistics in the plot corresponds to about a
  15 minute run or 50,000 decays. }
\label{LSD-MuDK}
\end{figure}

\section{Calibration strategy}

An important aspect of surface array detectors for UHECR studies is to
have a calibration strategy that allows to monitor the conversion of
the electronic signals into an equivalent energy deposit (or particle
count).

\subsection{Muon peak}
In the Auger surface array, the calibration of the WCD is based on the
energy deposited by a vertical muon traversing the WCD volume at its
centre~\cite{Bertou:2005ze,Ghia:2007fh}. The energy deposited by such
muons is about 240 MeV and is called a VEM (Vertical Equivalent
Muon). The constant flux of atmospheric muons (about 3000 go through
the Auger WCD each second) provides such a calibration. Indeed, 
the distribution of the charge collected at the dynode of any of the 3 PMT 
of the Auger WCD shows a hump (see figure~\ref{LSD-calib}) 
whose maximum corresponds nearly to one VEM  of energy deposited 
in the tank.  The exact correspondence between the VEM and the most probable charge 
collection has been measured in Auger using a scintillator telescope, and it is 
constant in time with only  a few percent variation over several years. 
Such a procedure gives the absolute calibration of individual PMTs.

For the absolute PMT calibration in the LSD a technique based on the
VEM value, similar to the one currently used in Auger, can be
adopted. An example for the bottom segment charge histogram of our
prototype (see next section for a detailed description) is shown in
figure~\ref{LSD-calib}.  From the ADC counts corresponding to the
maximum of the VEM charge histogram, one can find the correspondence
between ADC counts and the amount of energy deposited in the water
volume.

\subsection{Muon decay}
An alternative calibration can be obtained using the muon decays that
occur in the water volume. Nearly 4\% of the muons entering the WCD
stop and decay and the Michel electrons deposit on average an amount
of energy that can also be used for calibration purposes.

For the LSD it is also convenient to determine the water volumes
geometry, i.e. the precise height of the water separation interface
($h$) as the matrix coefficients $a$ and $b$ depend on it. The muon
decay rate in the top and bottom segments is determined by the
geometry of the station and hence allows for a precise determination
of the segment position $h$. 

In figure~\ref{LSD-MuDK} we show a simulation based on 50,000 muon
decays (this can be obtained in about 15 minutes assuming a decay
selection efficiency of 50\%) for three different positions of the
interface. A 5~cm difference corresponds to nearly 10 standard
deviations.
 
 A precision of a few millimeters can thus be achieved.

\subsection{Hybrids and physics data}
A cross calibration of the matrix coefficient within an array of LSD
is also possible based on physics results. Average (top and bottom)
LDFs from a set of events can be constructed (an example is given in
figure~\ref{LSD-avLDF}) for each individual LSD. From these average
LDFs and using the matrix one can reconstruct independently for each
LSD average muonic and EM LDFs. Since these average LDFs should be
identical for all LSDs this is a means to cross calibrate all matrices.

Finally in a hybrid observatory such as Auger~\cite{AugerNIM} or the
Telescope Array~\cite{Kawai:2008zza}, the EM LDFs from the LSD can be
calibrated using the fluorescence telescope data that give the cosmic
ray energy by means of a calorimetric measurement of the EM energy
deposit in the atmosphere~\cite{Verzi:2013eja,AbuZayyad:2012ru}. This
calibration scheme in the Auger Observatory uses the total signal in
the WCDs. In the LSD it will benefit from the EM signal reconstruction
as the uncertainty on the muon number in individual showers will no
longer deteriorate the energy resolutions of the surface array.

\subsection{Propagation of uncertainties and systematics}
When reconstructing the LDF from individual signals ($S$) measured in
standard WCDs the uncertainty of each measurements is of
the order of $\sqrt{S}$ with $S$ expressed in VEM units. This is due
to the Poisson fluctuation of the number of muons entering the tank
and to the fluctuation of the EM component in its high energy tail
which also introduce VEM size Poisson fluctuations. %
The additional uncertainty associated with the signal measurement in the
tank itself is due to the photo-electron (p.e.) statistics. As long as
we have number of p.e. per VEM much larger than one, this contribution to the error
budget is negligible.

Above 10~EeV, the signal at 1000~m is at least several tens of VEMs
and the particle fluctuations due to the detector sampling become
negligible compared to the shower to shower fluctuations.
ndeed above 10 EeV the energy resolution obtained from the LDF size
at 1000~m from the Auger WCD is 12\%~\cite{Abreu:2011pj} while one
would naively expect around 5 to 6\% if the particle count was alone
responsible for this uncertainty.
 
In an LSD tank the situation is similar even though the reconstructed
muon and EM signals are linear combinations of the top and bottom
signals with rather large coefficients.  For a matrix with
coefficients $a=0.6$ and $b=0.4$ we have :
\begin{equation}
 S_{EM}=3S_{top}-2S_{bot} \mbox{   and    }
 S_\mu=3S_{bot}-2S_{top}
 \label{smusem}  
 \end{equation}
However the particle fluctuations in the top and bottom segment are
correlated while the p.e. count fluctuations are not.  In the particular case of
the LSD geometry considered in this article, the contribution of the
p.e. count fluctuations in the reconstructed $S_\mu$ and $S_{EM}$ signals in each
tank, although amplified by the large coefficients of
equation~\ref{smusem}, is still less than 25\% of the particle
fluctuations themselves. This has little impact on the total error
budget, especially at the highest energies where shower to shower
fluctuations dominate.

Due to the relatively large coefficients entering in the
reconstruction of $S_\mu$ and $S_{EM}$ one must also consider the
effect of a possible systematic in the absolute calibration of the top
and bottom segments of the LSD. As an illustration, assume that
the calibration procedure induces a systematic bias of +1\% on the top
segment with respect to the bottom one over the whole array. In such case the EM signal from equation~\ref{smusem} will be on average 3\% too large and the
muon signal 2\% too small. Still, the LSD system can separate heavy
from light primaries according to the muonic content of the EAS as all
primaries will suffer the same systematic shift in the energy vs muon
size plane. The comparison with models becomes more difficult since
the electromagnetic size, hence the energy, will be overestimated
while the muon size will be underestimated. Given the calibration
strategy depicted above a maximum of 1\% systematic between the top
and bottom calibration is within reach but requires attention.

\begin{figure}[!t]
\center
\includegraphics[width=0.57\textwidth]{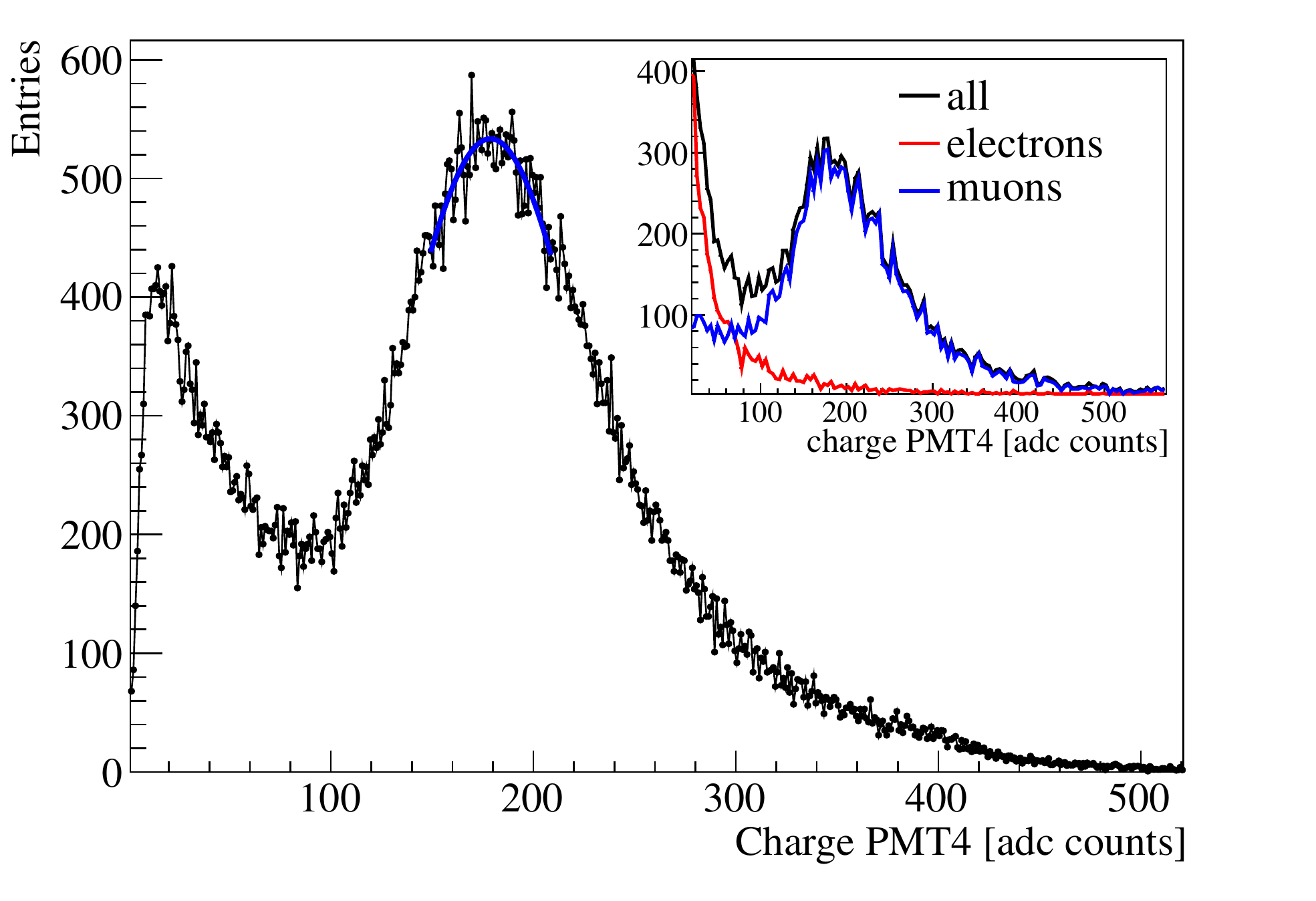}
\hfill
\includegraphics[width=0.41\textwidth]{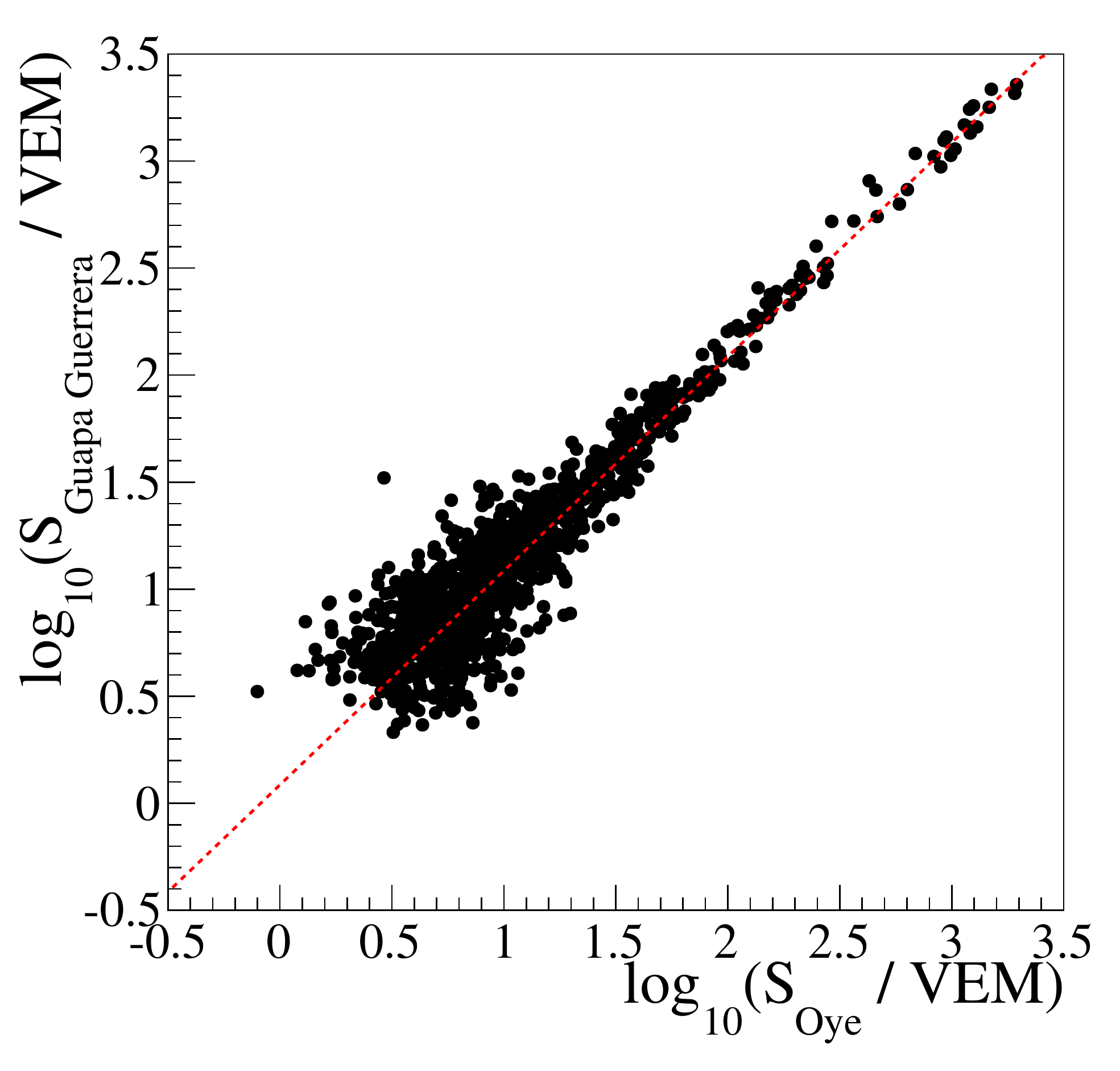}
\caption{\small Left: muon peak from the charge distribution of the
  bottom PMT of our prototype data compared with a GEANT4 simulations
  of a LSD (insert). Right: total signal in our LSD prototype (Guapa
  Guerrera) compared to a standard WCD (named Oye) of the Auger array located 10~m  away. As expected, there is a linear relation between the sum of the
  signals in the two LSD segments and a standard tank.}
\label{LSD-calib}
\end{figure}

\section{Prototype test}\label{sect:Prototype}

\begin{figure}[t]
\center
\hspace*{0ex}\includegraphics[width=0.49\textwidth]{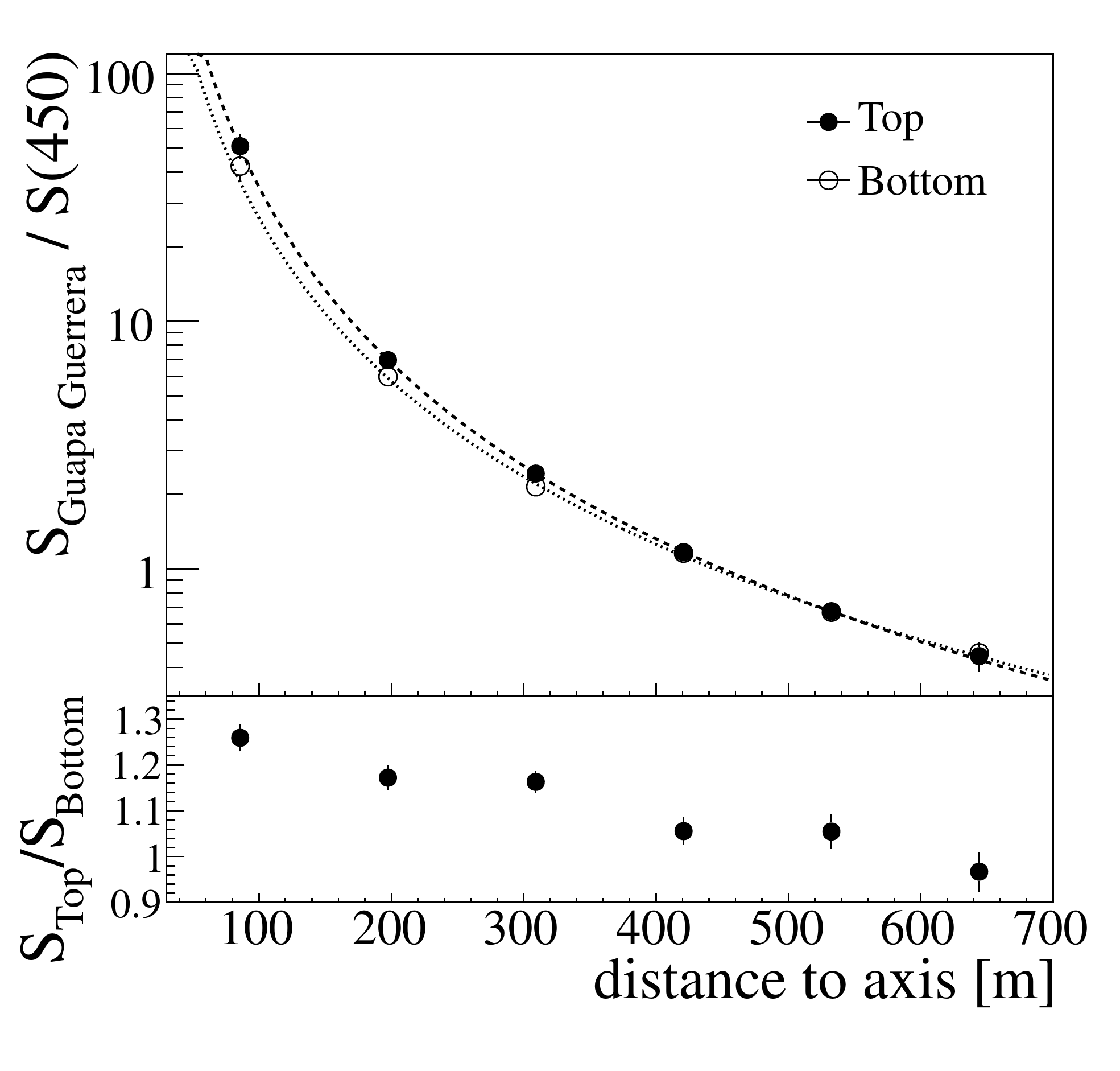}
\hfill
 \includegraphics[width=0.49\textwidth]{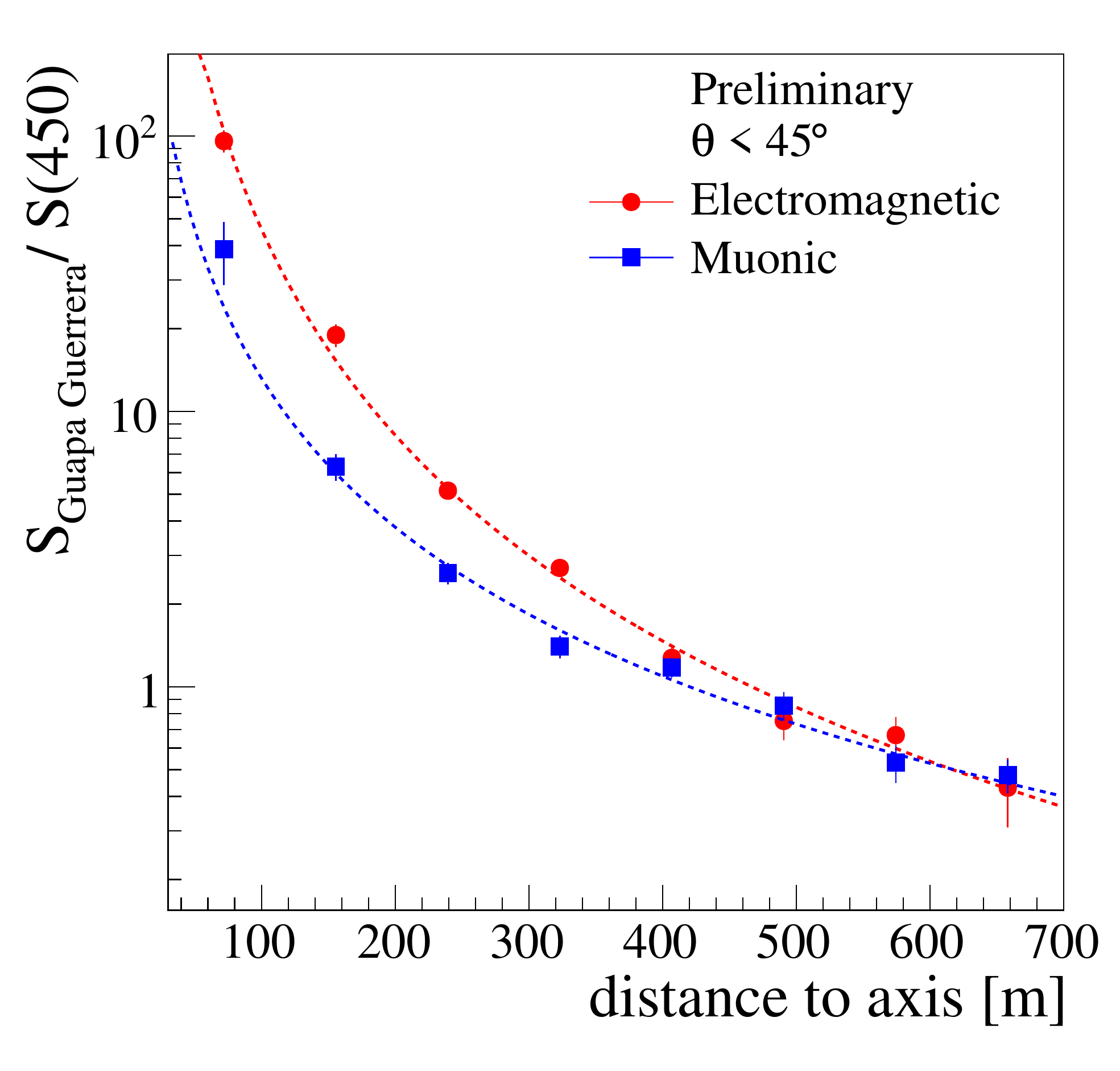}
\caption{\small Average LDF from the LSD prototype running at the
  Pierre Auger Observatory. 710 showers with reconstructed energies
  above 0.03 EeV, zenith angle below 45$^\circ$ and Guapa Guerrera
  located at a distance of less than 700~m have been selected for this
  plot. This corresponds to two weeks of data taking. Left: LDFs
  reconstructed with the signal recorded in the top and bottom layers
  separately. Right: preliminary muon and EM LDFs reconstructed
  applying a matrix with $a=0.6$ and $b=0.4$ to the top and bottom
  signals.}
\label{LSD-avLDF}
\end{figure}

A LSD prototype was constructed at the Auger site in Malarg\"ue
inserting a separation, made from a Tyvek laminate, in one of the Auger liners.  
This prototype uses the 4 PMT configuration displayed in figure~\ref{fig3}
with 3 PMTs looking into the top volume and 1 PMT looking in the
bottom one. This LSD, named ``Guapa Guerrera'', was installed in the
field during the last week of February 2014 and has been smoothly taking
data since then. We report here on the first two weeks of data collected by
this prototype.

\subsection{Calibration}

The histogram of charge of random signals for the PMT observing the
bottom volume, as recorded by the local acquisition
system~\cite{Bertou:2005ze}, is illustrated in Figure~\ref{LSD-calib},
left. From this figure one clearly distinguishes the peak which corresponds
to the charge deposited by single muons that is roughly one VEM (the
exact correspondence will be measured with a muon telescope) 
and can be used for calibration.

In the same figure, right, we have compared the total signal recorded
in the LSD (by summing after a preliminary calibration of top and
bottom signals) to the signal recorded in the same events by a
standard Auger WCD located 10~m away. We found, as expected, a linear
correlation between the two signals over nearly three orders of
magnitude. This shows that the LSD can also be used and can perform
like a standard WCD.

\subsection{Muonic and electromagnetic LDF reconstructions}
The LSD prototype is located in the infill part of the Auger surface
array, a region with a WCD density 4 times higher than in the regular
array~\cite{Ravignani:2013eja}.  It participates in about 100 physics
events per day. In the first two weeks of operation we recorded about
1400 events out of which we selected 710 events with a reconstructed
energy above 0.03 EeV and a zenith angle less than 45$^\circ$. We have
also requested that the energy deposit in the top part of the detector
is larger than $\approx$400 MeV. The zenith angle cut ensures that,
even for those relatively low energy showers that develop higher in
the atmosphere, the EM component is not completely absorbed before
reaching ground.  After normalizing the individual LDF of each events
at 450 m it is possible to plot, as a function of distance to cores,
the signal recorded by the Guapa Guerrera prototype in all of those
events.  This ``average LDF''
is shown on the left panel of figure~\ref{LSD-avLDF}
for the top and bottom segments separately. On can see from
this plot (and the accompanying lower panel showing the evolution of the top to bottom signals ratio)  
that the top average LDF is steeper than the bottom one. 
This is expected since the top segment has a larger contribution
from the EM component which has a steeper LDF than the muonic one.  On
the right panel of figure~\ref{LSD-avLDF} we show the reconstructed
average EM and muonic LDFs obtained from equation~\ref{smusem} applied
to the top and bottom average LDFs. This result is preliminary as we
used the nominal values of the matrix coefficients ($a=0.6$ and $b =
0.4$, not determined for the exact geometry of Guapa Guerrera) and
preliminary calibration constants.  Nevertheless the quality of this
result obtained after just two weeks of data taking is very
promising. In addition, and as stated previously, the reconstruction
of average muonic and EM LDFs using different LSD stations will allow us
to cross calibrate the individual matrices based on the requirement
that all LSDs are looking at the same physics.

\section{Conclusions}
We have presented the LSD (Layered Surface Detector) as a new concept
of water Cherenkov tank that allows us to reconstruct mass sensitive
parameters for UHECR with optimal resolution. The muon size of
EAS can be reconstructed with a precision better than 20\% above 10
EeV, reaching 10\% for energies above 70 EeV. The separation of muonic
and electromagnetic lateral distributions on an event by event basis
further provides an estimation of the \Xmax\, parameter with
50~g/cm$^2$ resolution from timing alone. It could be reduced to
30~g/cm$^2$ by including the time shape information.

A prototype of the LSD, constructed from a modification of one of Auger
surface array WCD, has shown excellent performances in agreement with
expectations from Monte-Carlo simulations.

We argue that such detectors should be seriously considered for any
upgrade of existing UHECR observatories or for the construction of new
observatories, either with larger aperture than the current one or
dedicated to the study of the second knee to ankle region, that is in
the energy range from 0.1~EeV to 10~EeV.

\section{Acknowledgments}
We gratefully acknowledge the very fruitful exchanges we had with all of our colleagues in the Auger collaboration and the use of all Auger facilities from hardware to software including access to a subset of shower data.  We are also
deeply indebted to the commitment of the observatory staff whose strong support in constructing and deploying the LSD prototype
was extremely appreciated.
The work of M.S. and I.C.M. made in the ILP LABEX (ANR-10-LABX-63), is supported by French state funds managed by the ANR
within the Investissements d'Avenir programme (ANR-11-IDEX-0004-02). I.C.M. also acknowledges the financial support by the
European Community 7th Framework Program, through the Marie Curie
Grant FP7-PEOPLE-2012-IEF, no. 328826.

\end{document}